 \documentclass[12pt,preprint]{aastex}

\def \apj {ApJ}
\def \apjs {ApJS}

\def \aap {A\&A}

\newcommand{\citeN}[1]{\citeauthor{#1} (\citeyear{#1})}
\newcommand{\citeNP}[1]{\citeauthor{#1} \citeyear{#1}}

\shortauthors{Centeno et al}
\shorttitle{Fine Structure in Chromospheric Umbral Oscillation}

\begin{document}

\title{Evidence for Fine Structure in the Chromospheric Umbral Oscillation}

\author{R. Centeno}
   	\affil{Instituto de Astrof\'\i sica de Canarias, Avda V\'\i a L\'
   	actea S/N, La Laguna 38205, Tenerife, Spain}
	\email{rce@iac.es}

\author{H. Socas-Navarro}
   	\affil{High Altitude Observatory, NCAR\thanks{The National Center
	for Atmospheric Research (NCAR) is sponsored by the National Science
	Foundation.}, 3450 Mitchell Lane, Boulder, CO 80307-3000, USA}
	\email{navarro@ucar.edu}

\author{M. Collados}
   	\affil{Instituto de Astrof\'\i sica de Canarias, Avda V\'\i a L\'
   	actea S/N, La Laguna 38205, Tenerife, Spain}
	\email{mcv@iac.es}

\author{J. Trujillo Bueno	\thanks{Consejo Superior de Investigaciones
    Cient\' \i ficas, Spain} 
} 
   	\affil{Instituto de Astrof\'\i sica de Canarias, Avda V\'\i a L\'
   	actea S/N, La Laguna 38205, Tenerife, Spain}
	\email{jtb@iac.es}

\date{}%

\begin{abstract}
Novel spectro-polarimetric observations of the \ion{He}{1} multiplet are used
to explore the dynamics of the chromospheric oscillation above sunspot
umbrae. The results presented here provide strong evidence in support of the
two-component model proposed by Socas-Navarro et al. According to this model,
the waves propagate only inside channels of sub-arcsecond width (the
``active'' component), whereas the rest of the umbra remains nearly at rest
(the ``quiet'' component). Although the observations support the fundamental
elements of that model, there is one particular aspect that is not compatible
with our data. We find that, contrarily to the scenario as originally
proposed, the active component remains through the entire oscillation cycle
and harbors both the upflowing and the downflowing phase of the oscillation.
\end{abstract}
   

\section{Introduction}
\label{intro}

The chromospheric oscillation above sunspot umbrae is a paradigmatic case of
wave propagation in a strongly magnetized plasma. This problem has drawn
considerable attention both from the theoretical (e.g., \citeNP{BHM+03}) and
the empirical standpoint (\citeNP{L92} and references therein). An
observational breakthrough was recently brought upon with the analysis of
spectro-polarimetric data, which is providing exciting new insights into the
process (\citeNP{SNTBRC00c}; \citeNP{SNTBRC00b}; \citeNP{SNTBRC01};
\citeNP{LASNM01}; \citeNP{CCTB05}). Previously to these works, the
chromospheric oscillation has been investigated by measuring the
Doppler shift of line cores in time series of intensity spectra. A particularly
good example, based on multi-wavelength observations, is the work of
\citeN{KMU81}.

Unfortunately, spectral diagnostics based on chromospheric lines is very
complicated due to non-LTE effects. Even direct measurements of chromospheric
line cores are often compromised by the appearance of emission reversals
associated with the upflowing phase of the oscillation, when the waves
develop into shocks as they propagate into the less dense chromosphere. A
very interesting exception to this rule is the \ion{He}{1} multiplet at
10830~\AA , which is formed over a very thin layer in the upper chromosphere
(\citeNP{AFL94})
and is not seen in emission at any time during the oscillation. These
reasons make it a very promising multiplet for the diagnostics of
chromospheric dynamics. However, there are two important observational
challenges. First, the long wavelength makes this spectral region almost
inaccessible to ordinary Si detectors (or only with a small quantum
efficiency). Second, the 10830 lines are very weak, especially the blue
transition which is barely visible in the spectra and is conspicuously
blended with a photospheric \ion{Ca}{1} line in sunspot umbrae.

In spite of those difficulties, there have been important investigations
based on 10830 Stokes~$I$ observations, starting with the pioneering work of
\citeN{L86}. More recently, the development of new infrared polarimeters
(\citeNP{RSL95}; \citeNP{CRHBR+99}; \citeNP{SNEP+04}) has sparked a renewed
interest in observations of the \ion{He}{1} multiplet, but now also with full
vector spectro-polarimetry. \citeN{CCTB05} demonstrated that the polarization
signals observed in the 10830 region provide a much clearer picture of the
oscillation than the intensity spectra alone. The sawtooth shape of the
wavefront crossing the chromosphere becomes particularly obvious in
fixed-slit Stokes~$V$ time series.

Before the work of \citeN{SNTBRC00c}, the chromospheric umbral oscillation
was thought of as a homogeneous process with horizontal scales of several
megameters, since these are the coherence scales of the observed
spectroscopic velocities. However, those authors observed the systematic
occurrence of anomalous polarization profiles during the upflowing phase of
the oscillation, which turn out to be conspicuous signatures of small-scale
mixture of two atmospheric components: an upward-propagating shock and a cool
quiet atmosphere similar to that of the slowly downflowing phase. This
small-scale mixture of the atmosphere cannot be detected in the intensity
spectra, but it becomes obvious in the polarization profiles because the
shocked component reverses the polarity of the Stokes~$V$ signal. The addition
of two opposite-sign profiles with very different Doppler shifts produces the
anomalous shapes reported in that work.

The results of \citeN{SNTBRC00c}, implying that the chromospheric shockwaves
have spatial scales smaller than $\sim$1'', still await independent
verification. In this work we looked for evidence in \ion{He}{1} 10830 data to
confirm or rebut their claim. It is important to emphasize that this
multiplet does not produce emission reversals in a hot shocked
atmosphere. Thus, one does not expect to observe anomalous profiles and the
two-component scenario would not be as immediately obvious in these
observations as it is in the \ion{Ca}{2} lines.

Here we report on a systematic study of the polarization signal in the
10830~\AA \, spectral region, with emphasis on the search for possible
signatures of the two-component scenario. We compared the observations with
relatively simple simulations of the Stokes profiles produced by one- and
two-component models. The results obtained provide strong evidence in favor
of the two-component scenario and constitute the first empirical verification
of fine structure in the umbral oscillation, using observations that
are very different from those in previous works.


\section{Observations}
\label{obs}

The observations presented here were carried out at the German Vacuum
Tower Telescope at the Observatorio del Teide on 1st October 2000, using
its instrument TIP (Tenerife Infrared Polarimeter, see \citeNP{MPCSA+99}),
which allows to take simultaneous images of the four Stokes 
parameters as a function of wavelength and position along the spectrograph
slit.
The slit was placed across the center of the umbra of a fairly regular spot, 
with heliographic coordinates 11S 2W,
and was kept fixed during the entire observing run ($\approx$ 1 hour).
In order to achieve a good signal-to-noise ratio, we added up
several images on-line, with a final temporal sampling of 7.9 seconds.

Image
stability was achieved by using a correlation tracker device (\citeNP{BCB+96}),
which compensates for the Earth's high frequency atmospheric 
variability, as well as for solar rotation.

The observed spectral range spanned from 10825.5 to 10833 \AA, with a spectral
sampling of 31 m\AA\ per pixel. 
This spectral region includes three interesting features: 
A photospheric Si {\sc i} line at 10827.09 \AA, a 
chromospheric Helium {\sc i} triplet  (around 10830 \AA), and a water vapour 
line (\citeNP{RSL95}) of telluric origin that can be used for 
calibration purposes, since it generates no polarization signal.



\clearpage
\begin{figure*}
\plotone{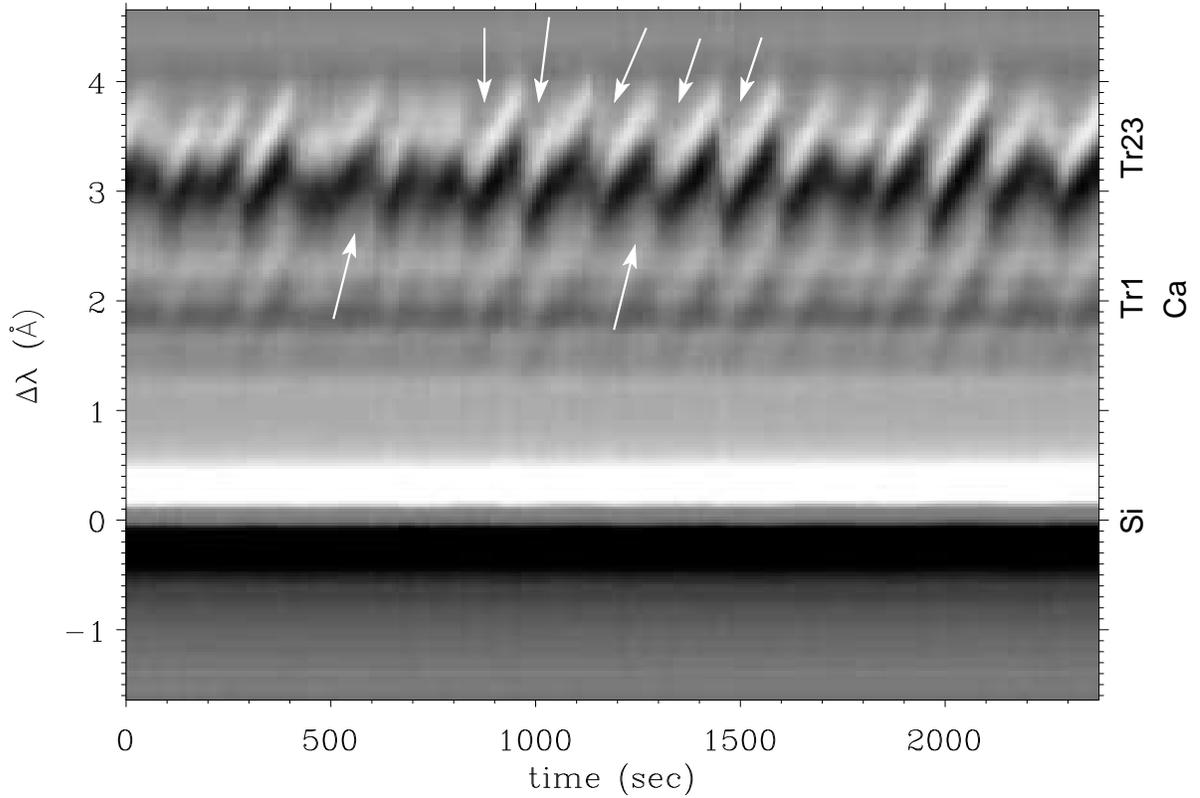}
\caption{Stokes~$V$ time series at a particular spatial position in the umbra
  of a sunspot. The labels on the right of the figure identify the spectral
  lines in the region shown. Wavelengths are measured in \AA \, from
  10827. The arrows indicate where the rest Tr23 component might be (barely)
  visible under the oscillating component (see discussion in the text).
\label{fig:series}
}
\end{figure*}
\clearpage

A standard reduction process was run over the raw data.
Flatfield and dark current measurements were performed at the beginning and
the end of the observing run and, in order to compensate for the telescope
instrumental polarization, we also took a series of polarimetric calibration
images. The calibration optics (Collados 1999) allows us to obtain the 
Mueller matrix of the light path between the instrumental calibration
sub-system and the detector. This process leaves a section of the telescope
without being calibrated, so further corrections of the residual cross-talk
among Stokes parameters were done: the $I$ to $Q$, $U$ and $V$ cross-talk was 
removed by forcing to zero the continuum polarization, and the circular
and linear polarization mutual
cross-talk was calculated by means of statistical techniques (Collados 2003).

\section{Interpretation}
\label{interp}


Let us first consider the relevant spectral features that produce significant
polarization signals in the 10830~\AA \, region. The He multiplet is
comprised of three different transitions, from a lower $^3S$ level with $J=0$
to three $^3P$ levels with $J=0, 1$ and~$2$. We hereafter refer to these as
transitions Tr1, Tr2 and Tr3 for abbreviation. Tr2 (10830.25 \AA) and 
Tr3 (10830.34 \AA) are blended and appear as one spectral line (Tr23) at 
solar temperatures. Tr1 (10829.09 \AA) is quite weak and relatively 
difficult to see in an intensity spectrum, while its Stokes $V$ signal
can be easily measured. A photospheric
\ion{Ca}{1} line is seen blended with Tr1, but is only present in the
relatively cool umbral atmosphere. Finally, a strong photospheric \ion{Si}{1}
line to the blue of Tr1 dominates the region both in the intensity and
the polarized spectra.


When the time series of Stokes~$V$ spectral images is displayed sequentially,
one obtains a movie that shows the oscillatory pattern of the He lines. In
the case of Tr1, the pattern is clearly superimposed on top of another
spectral feature that remains at rest. This feature has the appearance of a
broad spectral line with a core reversal similar to the magneto-optical
reversal observed in many visible lines. Fig~\ref{fig:series} shows the time
evolution of Stokes~$V$ at a particular spatial location in the umbra. The
figure shows the oscillatory pattern superimposed to the motionless feature
at the wavelength of Tr1 (we note that this is seen more clearly in the
movies). In our first analyses, at the beginning of this investigation, we
identified this feature at rest with the photospheric Ca line, which is not
entirely correct as we argue below. It is likely that other authors have made
the same assumption in previous works.
We would like to 
emphasize that this static spectral feature under
Tr1 is visible in all the temporal series of umbral oscillations 
we have so far (corresponding to different dates and different sunspots). 

Some of the arguments discussed in this section are based on a
Milne-Eddington simulation that contains all the transitions mentioned
above. In the simulation, the He lines are computed taking into account the
effects of 
incomplete Paschen-Back splitting (\citeNP{SNTBLdI04};
\citeNP{SNTBLdI05}). 

\subsection{Tr1}

Figure~\ref{fig:series} reveals the sawtooth shape of the chromospheric
oscillation in both Tr1 and Tr23, with a typical period of approximately
175~s . Every three minutes, the line profile undergoes a slow redshift
followed by a sudden blueshift (the latter corresponding to material
approaching the observer), resulting in the sawtooth shape
of the oscillation observed in Figure 1. This dynamical behavior evidences
shock wave formation at chromospheric heights. A detailed analysis is
presented in Centeno et al (2005).

Looking closely at Tr1, one can see what at
first sight would seem to be a photospheric umbral blend that does not move
significantly during the oscillation. A search over the NIST (National
Institute for Standards and Technology: http://www.physics.nist.gov) and VALD
spectral line databases (\citeNP{PKR+95}) produced only one possible match,
namely 
the umbral \ion{Ca}{1} line at 10829.27 \AA . We initially identified this
line with the blended feature because its wavelength, strength and umbral
character (i.e., it is only observed in susnpot umbrae) were in good
agreement with the data. However, when we tried to include this blend in our
Stokes synthesis/inversion codes, it became obvious that something was
missing in our picture.

\begin{figure*}
\plotone{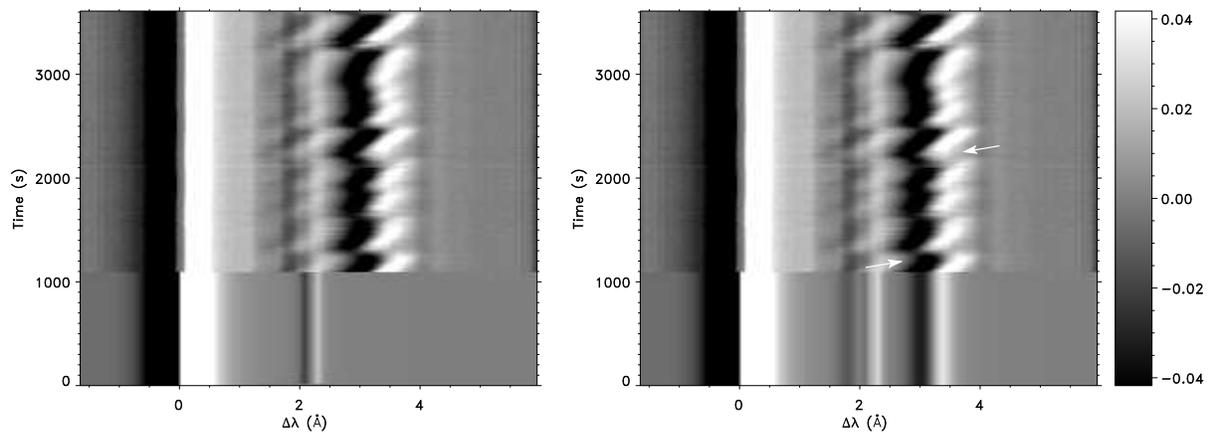}
\caption{ Similar to Figure 1, but the lower part of the image has been
  replaced by synthetic spectra. Left: Only the photospheric \ion{Si}{1} and
  \ion{Ca}{1} have been computed. Right: All four lines, including the
  \ion{He}{1} multiplet, are computed in a quiet component (zero
  velocity). Note how this simulation reproduces the observed spectral
  feature at 10829 \AA, including the core reversal. Wavelengths are measured
  in \AA \, from 10827. The arrows indicate where the rest Tr23 component
  might be (barely) visible under the oscillating component (see discussion
  in the text).
\label{fig:simser}
}
\end{figure*}

The left panel of Fig~\ref{fig:simser} represents a portion of an observed
time series at a fixed spatial point, with the lower part of the image
replaced with profiles of the Si and Ca lines produced by our
simulations. While the Si line appears to be correctly synthesized, the Ca
line clearly differs from the observed profile. There are three noteworthy
differences: a)The observed feature is much broader than the synthetic Ca
line. b)The synthetic profile does not appear to be centered at the right
wavelength. c)The observations exhibit a core reversal, very reminiscent of
the well-known magneto-optical effects that are sometimes seen in visible
lines.

We carried out more detailed simulations of the Ca line using an LTE code
(LILIA, \citeNP{SN01a}). We synthesized this line in variations of the
Harvard-Smithsonian reference atmosphere (HSRA, \citeNP{GNK+71}) and the
sunspot umbral model of \citeN{MAC+86}. These were used to look for
magneto-optical effects in the Stokes~$V$ profiles and also to verify the width
of the line shown in Fig~\ref{fig:simser} (left). We found that the LILIA
calculations confirmed the line width and none of them showed any signs of
Stokes~$V$ core reversals. Thus, the discrepancy between the simulations and
observations in the figure must be sought elsewhere.
\clearpage
\clearpage
The right panel of Fig~\ref{fig:simser} shows the same dataset, again with
the lower part replaced with a simulation. In this case the simulation
contains, in addition to the photospheric Si and Ca lines, the He multiplet
at rest. The synthesis was done with the incomplete Paschen-Back
Milne-Eddington code. Note how the combination of Tr1 with the Ca line
produces a spectral feature that is virtually identical to the observation,
with the correct wavelength, width and even the core reversal. This scenario,
with a quiet chromospheric component as proposed by \citeNP{SNTBRC00c},
naturally reproduces the observations with great fidelity. The core reversal
arises then as the overlapping of the blue lobe of the Ca profile with the
red lobe of Tr1.

\subsection{Tr2}

While Fig~\ref{fig:simser} presents a very convincing case in favor of the
two-component scenario, one would like to see the quiet component also under
the Tr23 line. Unfortunately, the simulations show that the quiet Tr23
profile must be obscured by the overlap with the active component. Only at
the times 
of maximum excursion in the oscillation it might be possible to observe a
brief glimpse of the hidden quiet component. One is tempted to recognize them
in both Figs~\ref{fig:series} and~\ref{fig:simser} (some examples are marked
with arrows). Unfortunately, such features are too weak to be sure.

One might also wonder why the quiet component is rather obvious under Tr1 but
not 
Tr23, since both lines form at approximately the same height and therefore
have the same Doppler width. However, Tr23 is broader since it is actually a
blend of two different transitions (Tr2 and Tr3) separated by $\sim$100~m\AA
. Under typical solar conditions, the Doppler width of Tr1 in velocity units
is $\sim$6~km~s$^{-1}$. The width of Tr23, taking into account the wavelength
separation of Tr2 and Tr3, is $\sim$9~km~s$^{-1}$. Comparing these values to
the amplitude of the chromospheric oscillation ($\sim$10~km~s$^{-1}$), we can
understand intuitively what the simulations already showed, namely that the
quiet component may be observed under Tr1 but only marginally (if at all)
under Tr23.


\section{Conclusions}
\label{conc}

The observations and numerical simulations presented in this work indicate
that the chromospheric umbral oscillation likely occurs on spatial
scales smaller than the resolution element of the observations, or
$\sim$1''. This suggests that the shock waves that drive the oscillation
propagate inside channels within the umbra\footnote{
Depending on the filling factor (which cannot be determined from these
observations alone due to the indetermination between theromdynamics and
filling factor), the scenario could be that we have small non-oscillating 
patches embedded in an oscillating umbra.
}. Recent
magneto-hydrodynamical simulations show that waves driven by a small piston
in the lower atmosphere remain confined within the same field lines as they
propagate upwards (\citeNP{BHM+03}). This means that photospheric or
subphotospheric small-scale structure is able to manifest itself in the
higher atmosphere, even if the magnetic field is perfectly homogeneous. 

The traditional scenario of the monolithic umbral oscillation, where the
entire chromosphere moves up and down coherently, cannot explain earlier
observations of the \ion{Ca}{2} infrared triplet made by
\citeN{SNTBRC00c}. Our results using the \ion{He}{1} multiplet support that
view in that the active oscillating component occupies only a certain filling
factor and coexists side by side with a quiet component that is nearly at
rest. However, our observations refute one of the smaller ingredients of the
model proposed by \citeN{SNTBRC00c}, namely the disappearance of the active
component after the upflow. In that work, the authors did not observe
downflows in the active component. For this reason, they proposed that the
oscillation proceeds as a series of jets that dump material into the upper
chromosphere and then disappear. In our data we can see the Tr1 and Tr23
lines moving up and down in the active component, which seems to indicate
that the active component remains intact during the entire oscillation
cycle. Other than that, the fundamental aspects of the two-component scenario
(i.e., the existence of channels in which the oscillation occurs), is
confirmed by the present work.

\acknowledgments 
This research has been partly funded by the Ministerio de Educaci\'on y
Ciencia through project AYA2004-05792 and by 
the European Solar Magnetism Network (contract HPRN-CT-2002-00313).


\end{document}